\documentclass[12pt,twoside]{article}
\usepackage[english]{babel}
\usepackage{amssymb}
\usepackage{graphicx}
\usepackage{epsfig}

\newcommand{\be}{\begin{equation}}
\newcommand{\ee}{\end{equation}}
\newcommand{\bea}{\begin{eqnarray}}
\newcommand{\eea}{\end{eqnarray}}
\newcommand{\ba}{\begin{array}}
\newcommand{\ea}{\end{array}}

\begin{document}

\begin{titlepage}

\begin{center}
\def\thefootnote{\fnsymbol{footnote}}
{\large The Kontsevich Connection on the Moduli Space of FZZT Liouville Branes}
\vskip 0.3in
S. Giusto \footnote{E-Mail: giusto@mps.ohio-state.edu} 
\vskip .2in
{\em Department of Physics, The Ohio State University\\
Columbus, OH 43210, USA}

\centerline{and} 
\vskip .2in
C. Imbimbo \footnote{E-Mail: imbimbo@ge.infn.it} 
\vskip .2in
{\em Dipartimento di Fisica, Universit\`a di Genova\\
and\\ Istituto Nazionale di Fisica Nucleare, Sezione di Genova\\
via Dodecaneso 33, I-16146, Genoa, Italy}
\end{center}
\vskip .2in
\begin{abstract}
We point out that insertions of matrix fields in (connected amputated)
amplitudes of (generalized) Kontsevich models are given by covariant
derivatives with respect to the Kontsevich moduli.  This implies that
correlators are sections of symmetric products of the (holomorphic)
tangent bundle on the (complexified) moduli space of FZZT Liouville
branes.  We discuss the relation of Kontsevich parametrization of
moduli space with that provided by either the $(p,1)$ or the $(1,p)$
boundary conformal field theories.  It turns out that the Kontsevich
connection captures the contribution of contact terms to open string
amplitudes of boundary cosmological constant operators in the $(1,p)$
minimal string models.  The curvature of the connection is of type
(1,1) and has delta-function singularities at the points in moduli
space where Kontsevich kinetic term vanishes.  We also outline the
extention of our formalism to the $c=1$ string at self-dual radius and
discuss the problems that have to be understood to reconciliate first
and second quantized approaches in this case.
\end{abstract}
\vfill
\end{titlepage}
\setcounter{footnote}{0}

\section{Introduction}
Kontsevich matrix models \cite{K,KMMMZ,AM,GN,IZ,IM} have been %
invented to compute correlators of observables in {\it closed} non-critical string theory
coupled to %
either  minimal or $c=1$ conformal matter. 
After the discovery of the AdS/CFT correspondence and the understanding of open/closed duality in the topological context  \cite{GV, DV}, it might have been natural to suspect 
that Kontsevich actions should be interpreted as %
 ``world volume'' actions for some  system of branes. Only a few years ago, however, 
advances in Liouville boundary conformal field theory \cite{FZZ,T} have 
allowed to identify precisely which branes are described by Kontsevich models: 
they are stable branes of Liouville theory coupled to minimal matter, also
known as FZZT branes. A derivation of the Kontsevich model as the Witten open
string field theory on these branes has been given recently \cite{GR} and only in the
case of $(1,2)$ matter --- which is equivalent to pure topological gravity.
So, curiously enough, the open string interpretation of Kontsevich models, 
which should represent, ``microscopically'',  their very definition, has been very
little explored: the present paper concentrates on this aspect.

For closed string computations, the observables of interest are 
$gl(N)$ invariant composite operators of the Kontsevich matrix field
 $X_{ij}$, with $i,j=1,\ldots, N$.  The physics of open string stretching between $N$ branes 
is captured instead by ``colored''  correlators of the elementary field $X_{ij}$. $X$ is the string field corresponding to a primary vertex operator of the boundary conformal field theory, that, as we will
see, is the boundary cosmological constant operator $\mathcal{B}$. 
$\mathcal{B}$ is a marginal operator and this manifests itself as the fact that the Kontsevich action also depends on %
an external matrix source $Z_{ij}$. The eigenvalues of $Z$ are the variables by which 
the string field theory parametrizes the moduli space of FZZT branes. Naively, insertions of 
$\mathcal{B}$  in world-sheet %
correlators  are given by derivatives with respect to the moduli $Z$.
One of the lessons from topological string theories \cite{BCOV} is that in general
this is true %
only up to contact terms, which in our case come from the points in moduli space 
where open vertex operators collide between themselves or with 
a node of the surface. The contribution of contact terms is to convert simple
derivatives with respect to the moduli into covariant ones.
In this paper we show that this is exactly what happens
also in the case of the Kontsevich model:  the connected amputated $n$-point 
correlator of the matrix fields  $X$ is given by the $Z_{ij}$-covariant derivative 
of the $n-1$-point correlator. We deduce the connection which defines the 
covariant derivative from a differential equation satisfied by the effective
potential of the model, as a function of the matrix  $Z$ and of a matrix field source $\phi$. 
We do this in Section 2 for the case of the 
$(1,2)$ model and in Section 4 for the general $(1,p)$ models and $c=1$.
Extending the matrix $Z$ to the complex domain, one finds that 
the curvature of this connection has  vanishing $(2,0)$ part, as required 
by the symmetry of the correlators with
respect to exchange of two fields $X$. 
The $(1,1)$ part of the curvature, however, is non-zero
and has delta-functions with support at the points in $Z$ moduli space
at which the Kontsevich kinetic term vanishes. At these
points the Kontsevich matrix theory becomes singular, %
perhaps signaling the fact that more degrees
of freedom are needed to describe the physics 
in these regions of the moduli space.  Such a possibility is not %
unlikely since the Kontsevich matrix models are reductions of the full-fledged Witten open string
field theory and should be thought of as effective theories. 

From the point of view of the boundary conformal field theory, the computation
of contact terms is generally a difficult problem. The Kontsevich connection, 
encoding the information on these contacts, thus provides predictions that are quite non-trivial. 
In Section 3, a comparison between the Kontsevich results and 
the world-sheet computations is carried out, in the case
of a single brane and at string tree level. This comparison, apart from 
elucidating some conceptual issues regarding the consistency between string
field theory and conformal field theory descriptions, also provides 
non-trivial support for the conjecture, not yet explicitely proven 
beyond $p=2$, that
Kontsevich theory for $(1,p)$ minimal strings is the (effective) string
field theory of FZZT branes coupled to minimal matter.

In order to compare with boundary conformal field theory results, one first
has to understand the relation between the string field theory parameter
$z$ and the conformal field theory modulus, which, in our case, is the 
boundary cosmological constant $\mu_B$. In general \cite{SZ,S},
string field theory moduli and conformal field theory moduli
give different parametrization of the open string moduli space. 
In this context,  we find it convenient  to relate our analysis to the recent results presented 
in  \cite{SS}, which have shown that disk string 
amplitudes of boundary cosmological operators in
$(p,q)$ minimal models are encoded in their closed ground rings.  
Since the closed ground ring equations also naturally appear in 
Kontsevich models as the equations of motion, the picture of \cite{SS} is the
natural one to prove the equivalence between Kontsevich open string field
theory and open minimal strings. The upshot of the discussion of 
Section 3 is the following: The Kontsevich parameter $z$ is a non-trivial
function of the boundary cosmological constant of the $(p,1)$ theory --- it is the
dual boundary cosmological constant. Correlators of $n$ boundary operators have to be 
interpreted as sections of the $n$-th symmetric product of the tangent space 
to the moduli space. The Kontsevich connection defines a covariant derivative on this bundle.
This connection vanishes in the parametrization of moduli space associated to the conformal
$(p,1)$ model but not in the parametrization  associated %
to the Kontsevich model.  
The reason why this is possible is that Kontsevich 
theory, unlike the conformal  $(p,1)$ theory, picks up the {\it globally}
defined parameter on the moduli space. Thus, the example we are considering
confirms the belief \cite{SZ,S} that open string field theory always provides a one-to-one
parametrization  of moduli space.  Since there is a non-trivial reparametrization
between the moduli of the open string field theory and  those of the boundary conformal field theory, correlators computed in the two theories are  equal only 
up to a coordinate transformation on moduli space. 

The existence of a non-trivial Kontsevich connection also elucidates
the equivalence of the string theories based on the $(p,1)$ and $(1,p)$ conformal models.
The parametrization of moduli space of the $(1,p)$ model is the globally
defined one: therefore, the corresponding connection does not vanish,
unlike the one of the $(p,1)$ model. This means that amplitudes of the $(1,p)$ model
must receive non-trivial contributions from contact terms:  it is  only by  taking these
into account that one can establish the equality --- up to reparametrization --- of these amplitudes
with those of the $(p,1)$ model.

Finally, in Section 4.2, we extend the analysis to the Kontsevich model
 \cite{IM} for the $c=1$ non-critical strings at self-dual radius and we conjecture that 
also in this case the matrix field  correlators, that we compute, should be identified with amplitudes of
boundary cosmological constant operators. In this case, however, a 
comparison with 
worldsheet computations appears at present problematic: The naive limit of 
correlators of Liouville boundary cosmological
constant operators for $c=1$ is divergent and no univocal prescription 
to ``renormalize''  these divergences has been understood so far.

\section{The Kontsevich Connection}
(Generalized) Kontsevich matrix models are defined by the following
matrix integral
\be
{\rm e}^{F(g_s, Z)}= \int\! [dM] \, 
{\rm e}^{-{1\over g_s}{S_K}(M,Z)}
\label{kfree}
\ee 
where both $M$ and $Z$ are hermitian $N\times N$ matrices and the action
\be
S_K(M, Z) \equiv {\rm Tr} \bigl[V(M) -V(Z) - 
V'(Z) (M-Z)\bigr]
\label{kaction}
\ee
depends on the potential $V(M)$. 

In the early nineties it was understood that 
when the potential $V(M)$ is the polynomial
\be
V(M) = {M^{q+1}\over q+1}
\label{qpotential}
\ee
with $q$ integer greater than 1, the associated Kontsevich matrix
model (\ref{kfree}) is related to {\it closed} bosonic string
theory coupled to minimal conformal matter of the $(p,q)$ type.
Kontsevich originally considered the model 
with $q=2$: 
\be
{\rm e}^{F_2(g_s, Z)}= {\rm e}^{-{2\,\over 3\, g_s} {\rm Tr}\,Z^3}\int\! [dM] \, 
{\rm e}^{-{1\over g_s}{\rm Tr} \bigl[{M^3\over 3} -Z^2\,M\bigr]}=
\int\! [dX] \, 
{\rm e}^{-{1\over g_s}{\rm Tr} \bigl[Z\,X^2+ {X^3\over 3}\bigr]}
\label{kintegral}
\ee 
where we have put
\be
M = Z +X
\label{shift}
\ee
He showed that the matrix model (\ref{kintegral}) encodes the correlators
at all genus of pure
topological gravity. This latter theory had been understood 
to be equivalent to 
the double scaled one-matrix models
which describe bosonic closed strings coupled to $(p,2)$ conformal matter. 

The way in which the model (\ref{kintegral}) computes the correlators
of topological gravity is the following: Let us denote by $O_n$, 
with $n=1,2,\ldots$ the  observables of closed topological gravity. 
The correlators of the $O_n$'s
are encoded in the generating function 
\be
Z_{top. grav.} (g_s, t_n) \equiv \langle {\rm e}^{\sum_n t_n\, O_n}\rangle
\label{ztopgrav}
\ee
which depends on both the closed string coupling constant $g_s$ and 
the infinite number of variables $t_n$. Let also $z_i$, with $i=1,\ldots
,N$, be the $N$ eigenvalues of the hermitian matrix $Z$. Kontsevich showed 
that
\be
Z_{top. grav.} (g_s, t_n)\Big|_{t_n =t_n (z_i)} = {\cal N}(Z)^{-1}\,{\rm e}^{F_2(g_s, Z)}
\label{openclosed}
\ee
where the functions $t_n(z_i)$ are defined by means of the
the Frobenius-Miwa-Kontsevich transform
\footnote{Standard arguments of matrix field theory show that the
free energy $F_2(g_s, Z)$ on the R.H.S. of Eq. (\ref{openclosed})
does indeed depend on $g_s$ and the combinations (\ref{fmk})
of the eigenvalues $z_i$.}  
\be
t_n (z_i) \equiv -{g_s \over n}\,{\rm Tr}\, Z^{-n}
\label{fmk}
\ee
and the normalization factor ${\cal N}(Z)$ is given by the quadratic
matrix integral
\be
{\cal N}(Z)=
\int\! [dX] \, 
{\rm e}^{-{1\over g_s}{\rm Tr}\,Z\,X^2}
\label{normalization}
\ee 

Recently it was proposed \cite{GR} that Kontsevich matrix integral 
(\ref{kintegral}) be interpreted as Witten open string field theory
on $N$ stable branes of non-critical bosonic strings in the 
background of (1,2) conformal minimal matter. 
These branes have boundary conditions
of the FZZT type along the Liouville direction. FZZT boundary conditions
are parametrized by the {\it boundary} cosmological constant $\mu_B$.
The  parameter $\mu_B$ shows up in the  Liouville action on the 
world-sheet $\Sigma$ with boundary $\partial \Sigma$ as the coupling
constant that multiplies the boundary cosmological constant vertex operator
$\mathcal{B}_b(\phi)\equiv {\rm e}^{b\,\phi}$
\be
S_{boundary} = \mu_B\int_{\partial \Sigma}\!\!\!{\rm e}^{b\,\phi}
\label{boundaryaction}
\ee
where $\phi$ is the Liouville field, and the parameter $b$ is related
to the Liouville Virasoro central charge
\be
c_{Liouville} = 1+ 6\, Q^2 \qquad Q= b + {1\over b}
\ee
When the matter is represented by the $(p,q)$ minimal models, 
\be
b^2 = {p\over q}
\ee
According to \cite{GR}, the matrix $X$ in the Kontsevich
integral (\ref{kintegral}) is to be identified with 
the second quantized field corresponding
to the boundary cosmological constant vertex operator $\mathcal{B}_b(\phi)$. 
The eigenvalues $z_i$ of the
matrix $Z$ which appears in the Kontsevich action
define coordinates on the moduli space of FZZT branes. In the boundary
conformal field theory the same moduli space is parametrized by the boundary 
cosmological constants $\mu_B^{(i)}$ on the $i$-th brane. In general 
\cite{SZ,S}, open string field theory and conformal field theory coordinates 
are non-trivially related. In \cite{GR}, by comparing Kontsevich and
FZZT amplitudes for the $(1,2)$ theory, it was found that this relation is
simply given by
\be
\mu_B^{(i)} =z_i
\label{ztomub}
\ee
Since, however, the $(1,2)$ and $(2,1)$ theory are physically equivalent,
Kontsevich field theory must provide predictions which are consistent with
both.  
In the following we explore the implications of the
identifications of \cite{GR} for the string amplitudes
of $n$ boundary cosmological constant operators $\mathcal{B}_b(\phi)$: We will 
explain, among other things, how the $(1,2)$ and $(2,1)$ model are related
and in which way Kontsevich theory gives a consistent description of both. 

Connected correlators of $n$ matrix fields $X_{ij}$
\be
\langle X_{i_1j_1}\cdots X_{i_nj_n}\rangle
\ee
are given by
differentiating $n$ times the generating function $F_X(g_s, Z,J)$ 
\be
{\rm e}^{F_X(g_s, Z,J)}\equiv \int\! [dX] \, 
{\rm e}^{-{1\over g_s}{\rm Tr} \bigl[Z\,X^2+ {X^3\over 3}\bigr]+ 
{\rm Tr}\, J\, X}
\label{kconnected}
\ee
with respect to the matrix source $J_{ji}$ and then setting $J$ to zero. 
(Connected) string amplitudes equal {\it amputated} connected correlators
of the second quantized theory.
Therefore, string amplitudes with $n$ boundary cosmological integrated
vertex operators $\mathcal{B}_b(\phi)$ are given by differentiating $n$ times 
the function 
\be
F_X(g_s, Z, {1\over g_s}\{Z,\phi\})
\label{kamputated}
\ee
with respect to $\phi_{ji}$ and then setting $\phi$ to zero. 
It is useful to introduce the effective potential $H(Z, \phi)$ which is
defined by subtracting to the generating function of the
{\it amputated} connected correlators its free part:
\be
H(Z,\phi)\equiv F_X(g_s, Z, {1\over g_s}\{Z,\phi\})-\log {\cal N}(Z) -
{1\over g_s}{\rm Tr}\, Z\,\phi^2
\label{effectivepotential}
\ee

We now want to derive an equation that allows to express insertions of
fields $X$ in amputated connected amplitudes in terms of derivatives
with respect to $Z$. The existence of such an equation might be expected
a priori, given the relation (\ref{ztomub}) and the fact
that insertions of integrated vertex operators $\mathcal{B}_b(\phi)$ in string
amplitudes should be given --- up to contact terms ---
by derivatives with respect to $\mu_B$ (see Eq. (\ref{boundaryaction})). 
We will see that Kontsevich theory implies specific values of the
contact terms.

To derive this equation, let us go back to the integration
variable $M$ in Eq. (\ref{kconnected}):
\bea
&&{\rm e}^{F_X(g_s, Z,J)}=
 {\rm e}^{-{2\,\over 3\, g_s} {\rm Tr}\,Z^3-{\rm Tr}\,J\,Z}
\int\! [dM] \,{\rm e}^{-{1\over g_s}{\rm Tr} 
\bigl[{M^3\over 3} -(Z^2+g_s J)\,M\bigr]}=\nonumber\\
&&\qquad\qquad ={\rm e}^{-{2\,\over 3\, g_s} {\rm Tr}\,Z^3-{\rm Tr}\,J\,Z+ 
F_2(g_s,(Z^2+g_s J)^{1\over2})+{2\,\over 3\, g_s} {\rm Tr}\,
(Z^2+g_s J)^{3\over2}}
\label{kconnectedbis}
\eea
and thus
\bea
&&H(Z,\phi) = F_2(g_s, \bigl[Z^2 +\{Z,\phi\}\bigr]^{1\over 2})-\log {\cal N}(Z)+\nonumber\\
&&\quad+ {2\,\over 3\, g_s}\bigl[{\rm Tr}\,
(Z^2+\{Z,\phi\})^{3\over2}-{\rm Tr}\,Z^3 -
3\, {\rm Tr}\,Z^2\,\phi -{3\over 2 }{\rm Tr}\, Z\,\phi^2\bigr]=\nonumber\\
&&\quad = f_2(Y)-\log {\cal N}(Z)- {2\,\over 3\, g_s}\bigl[{\rm Tr}\,Z^3 +
3\, {\rm Tr}\,Z^2\,\phi +{3\over 2 }{\rm Tr}\, Z\,\phi^2\bigr]
\label{effectivepotentialbis}
\eea
where we introduced the hermitian $N\times N$ matrix
\be
Y \equiv Z^2 +\{Z,\phi\}
\label{deformedZ}
\ee
and the invariant function $f_2(Y)$ of $Y$
\be
f_2(Y) \equiv F_2(g_s, Y^{1\over 2}) +{2\,\over 3\, g_s}{\rm Tr}\,Y^{3\over2}
\ee
It follows from Eq. (\ref{kintegral}) that $f_2(Y)$ satisfies
\be
{\rm e}^{f_2(g_s, Y)} = \int\! [dM] \,{\rm e}^{-{1\over g_s}{\rm Tr} 
\bigl[{M^3\over 3} -Y\,M\bigr]}
\label{measure}
\ee
Eq. (\ref{effectivepotentialbis}) essentially states that the generating function
$H(Z,\phi)$ depends on $Z$ and $\phi$ only through the combination $Y$. In the following
we want to translate this property of $H(Z,\phi)$ into a differential equation.

From Eq. (\ref{deformedZ}) one obtains
\be
{\partial Y_{ij}\over \partial Z_{kl}} = {\partial Y_{ij}\over \partial 
\phi_{kl}}+ \delta_{ik}\,\phi_{lj} + \phi_{ik}\,\delta_{lj}
\ee
Therefore
\be
{\partial f_2(Y)\over \partial Z_{kl}} = 
{\partial f_2(Y)\over \partial \phi_{kl}}+ \{\phi,{\partial f_2(Y)\over \partial Y}\}_{lk}
\label{zversusphi}
\ee
Moreover
\be
{\partial f_2(Y)\over \partial \phi_{kl}}=\sum_{i,j}
\bigl(Z _{ik}\,\delta_{lj} + \delta_{ik} Z_{lj}\bigr)
{\partial f_2(Y)\over \partial Y_{ij}} \equiv \sum_{i,j}\Delta_{lk;ji}(Z){\partial f_2(Y)\over \partial Y_{ij}}
\label{ampversuscon}
\ee
where we introduced $\Delta_{lk;ji}(Z)$, the inverse of the propagator of
Kontsevich matrix theory. 
Going to the basis in which $Z$ is diagonal,
\be
\Delta_{lk;ji}^{-1}(Z) = {\delta_{lj}\,\delta_{ki}\over z_i + z_j}
\label{kpropagator}
\ee
Therefore, in such basis one has
\be
{\partial f_2(Y)\over \partial Y_{kl}} ={1\over z_k + z_l}{\partial f_2(Y)\over \partial \phi_{kl}}
\ee
Plugging this into (\ref{zversusphi}) we obtain
\be
{\partial f_2(Y)\over \partial Z_{kl}} = 
{\partial f_2(Y)\over \partial \phi_{kl}}+ \sum_i{\phi_{li}\over z_i+z_k}\,{\partial f_2(Y)\over \partial \phi_{ki}} +  {\phi_{ik}\over z_i+z_l}\,
{\partial f_2(Y)\over \partial \phi_{il}}
\label{zversusphibis}
\ee
Hence, from (\ref{effectivepotentialbis})  we deduce
\bea
&&\Bigl[{\partial\over \partial Z_{kl}} - {\partial\over \partial \phi_{kl}}-\sum_i{\phi_{li}\over z_i+z_k}\,{\partial\over \partial\phi_{ki}} -  
{\phi_{ik}\over z_i+z_l}\,{\partial\over \partial \phi_{il}}\Bigr] H(Z,\phi)=\nonumber\\
&&\quad=-\Bigl[{\partial\over\partial Z_{kl}}-{\partial\over \partial \phi_{kl}}-\sum_i{\phi_{li}\over z_i+z_k}\,{\partial\over \partial\phi_{ki}} -  
{\phi_{ik}\over z_i+z_l}\,{\partial\over \partial \phi_{il}}\Bigr]\nonumber\\
&&\Bigl[\log {\cal N}(Z)+{2\,\over 3\, g_s}\bigl[{\rm Tr}\,Z^3 +
3\,{\rm Tr}\,Z^2\,\phi+{3\over 2 }{\rm Tr}\, Z\,\phi^2\bigr]\Bigr]
\label{derivativeZ}
\eea
Since
\be
{\partial\over\partial Z_{kl}}{2\,\over 3}\bigl[{\rm Tr}\,Z^3 +
3\,{\rm Tr}\,Z^2\,\phi+{3\over 2 }{\rm Tr}\, Z\,\phi^2\bigr] =
\bigl(2\,Z^2 + 2\,\{Z,\phi\} +\phi^2\bigr)_{lk}
\ee
and
\be
{\partial\over\partial \phi_{kl}}{2\,\over 3}\bigl[{\rm Tr}\,Z^3 +
3\,{\rm Tr}\,Z^2\,\phi+{3\over 2 }{\rm Tr}\, Z\,\phi^2\bigr] =
\bigl(2\,Z^2 + \{Z,\phi\}\bigr)_{lk}
\ee
it follows that
\bea
&&\Bigl[{\partial\over\partial Z_{kl}}-{\partial\over \partial \phi_{kl}}-\sum_i{\phi_{li}\over z_i+z_k}\,{\partial\over \partial\phi_{ki}} -  
{\phi_{ik}\over z_i+z_l}\,{\partial\over \partial \phi_{il}}\Bigr]{2\,\over 3\, g_s}\bigl[{\rm Tr}\,Z^3 +\nonumber\\
&&\qquad\qquad
+3\,{\rm Tr}\,Z^2\,\phi+{3\over 2 }{\rm Tr}\, Z\,\phi^2\bigr]= -{1\over g_s}\,(\phi^2)_{lk}
\eea
Moreover
\be
{\partial\log {\cal N}(Z)\over\partial Z_{kl}} =-\delta_{lk}\sum_i {1\over z_i+z_k}
\ee
Eq.({\ref{derivativeZ}) becomes therefore
\bea
&&\Bigl[{\partial\over \partial Z_{kl}} - {\partial\over \partial \phi_{kl}}-\sum_i{\phi_{li}\over z_i+z_k}\,{\partial\over \partial\phi_{ki}} -  
{\phi_{ik}\over z_i+z_l}\,{\partial\over \partial \phi_{il}}\Bigr] H(Z,\phi)=\nonumber\\
&&\quad= \delta_{lk}\sum_i {1\over z_i+z_k}+{1\over g_s}\,(\phi^2)_{lk} 
\label{derivativeZbis}
\eea
Introducing the generating function of the (both connected and non connected) amputated
correlators of $X$  
\be
{\cal Z}(Z,\phi) \equiv {\rm e}^{H(Z,\phi)}
\ee
we obtain the homogeneous linear equation 
\bea
&&\Bigl[{\partial\over \partial Z_{kl}} - {\partial\over \partial \phi_{kl}}-\sum_i{\phi_{li}\over z_i+z_k}\,{\partial\over \partial\phi_{ki}} -  
{\phi_{ik}\over z_i+z_l}\,{\partial\over \partial \phi_{il}} +\nonumber\\
&&\qquad\quad -\delta_{lk}\sum_i {1\over z_i+z_k}-
{1\over g_s}\,(\phi^2)_{lk}\Bigr] {\cal Z}(Z,\phi) =0
\label{holoequation}
\eea

Eq.(\ref{derivativeZbis}) relates $X$ insertions in correlators to 
derivatives with respect to $Z_{kl}$. Let us see explicitly how this works. 
Let us introduce the covariant $\phi$-derivative
\be
D_{lk}^{\phi}\equiv {\partial\over \partial \phi_{kl}}+
\sum_i \bigl({\phi_{li}\over z_i+z_k}\,{\partial\over \partial\phi_{ki}} +  
{\phi_{ik}\over z_i+z_l}\,{\partial\over \partial \phi_{il}}\bigr)
\label{phiderivative}
\ee
Since 
\be
\Bigl[{\partial\over\partial \phi_{mn}},D_{lk}^{\phi}\Bigr]=
{\delta_{lm}\over z_n+z_k}\,{\partial\over \partial\phi_{kn}}+
{\delta_{nk}\over z_m+z_l}\,{\partial\over \partial \phi_{ml}}
\ee
derivation of Eq. (\ref{derivativeZbis}) with respect $\phi_{mn}$
gives
\bea
&&\Bigl[{\partial\over \partial Z_{kl}}
{\partial H(Z,\phi)\over \partial\phi_{mn}} - {\delta_{lm}\over z_n+z_k}\,
{\partial H(Z,\phi)\over \partial\phi_{kn}}-
{\delta_{nk}\over z_m+z_l}\,{\partial H(Z,\phi)\over \partial \phi_{ml}}\Bigr]
=\nonumber\\
&&\quad=D_{lk}^{\phi} {\partial H(Z,\phi)\over \partial \phi_{mn}}
+{1\over g_s}\,(\delta_{lm}\phi_{nk}+\phi_{lm}\delta_{nk}) 
\label{recursion}
\eea
Setting now $\phi_{ij}=0$, one expresses the connected amputated 2-point
function of $X$ as $Z$ covariant derivative of the 1-point function:
\bea
&&{\;\partial^2 H(Z,\phi)\;\over \partial \phi_{kl}\, \partial \phi_{mn}}=
\langle X_{lk}\,X_{nm}\rangle_{c.a.}-{1\over g_s}(\delta_{nk} Z_{lm}+Z_{nk}\delta_{lm}) 
=\nonumber\\
&&\qquad\qquad= {\partial\,\langle X_{nm}\rangle_{c.a.}\over \partial Z_{kl}} 
-{\delta_{lm}\langle X_{nk}\rangle_{c.a.}\over z_n+z_k}\,-
{\delta_{nk}\langle X_{lm}\rangle_{c.a.}\over z_m+z_l}=\nonumber\\
&&\qquad\qquad = {\partial\,\langle X_{nm}\rangle_{c.a.}\over \partial Z_{kl}}
- \sum_{i,j} {\delta_{lm}\,\delta_{ki}\,\delta_{nj}+\delta_{nk}\,\delta_{mi}\,
\delta_{lj}
\over z_i+z_j}\,\langle X_{ji}\rangle_{c.a.}=\nonumber\\
&&\qquad\qquad \equiv{D\over D\, Z_{kl}}\langle X_{nm}\rangle_{c.a.}
\label{twopoints}
\eea
The two terms that make up the connection piece of the covariant
derivative ${D\over D\, Z_{kl}}$ have a clear physical origin:
they are the two contact terms that are generated when
the operator $X_{lk}$ collides with the operator $X_{nm}$ either
from the left or from the right (see Figure~\ref{f:contacts.eps}).
\begin{figure}
\begin{center}
\includegraphics*[scale=.6, clip=false]{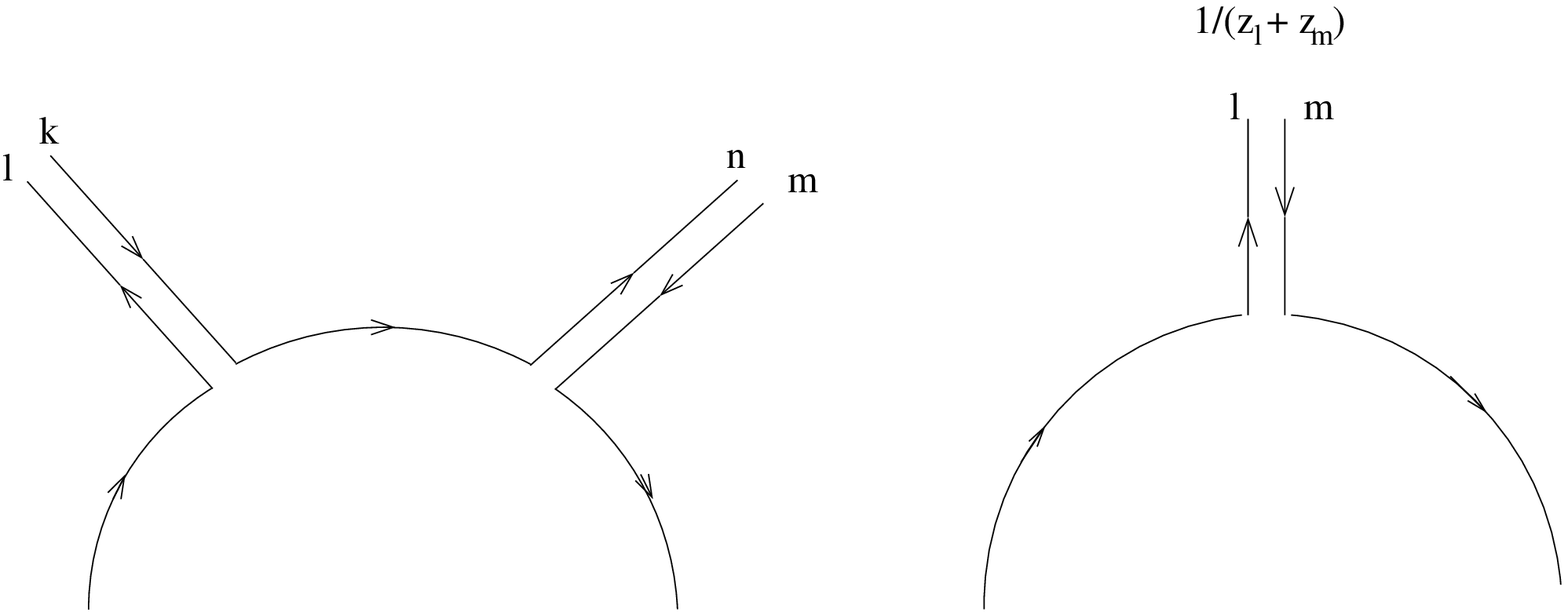}
{$\centerline{\rm (a)}$}
\bigskip
\includegraphics*[scale=.6, clip=false]{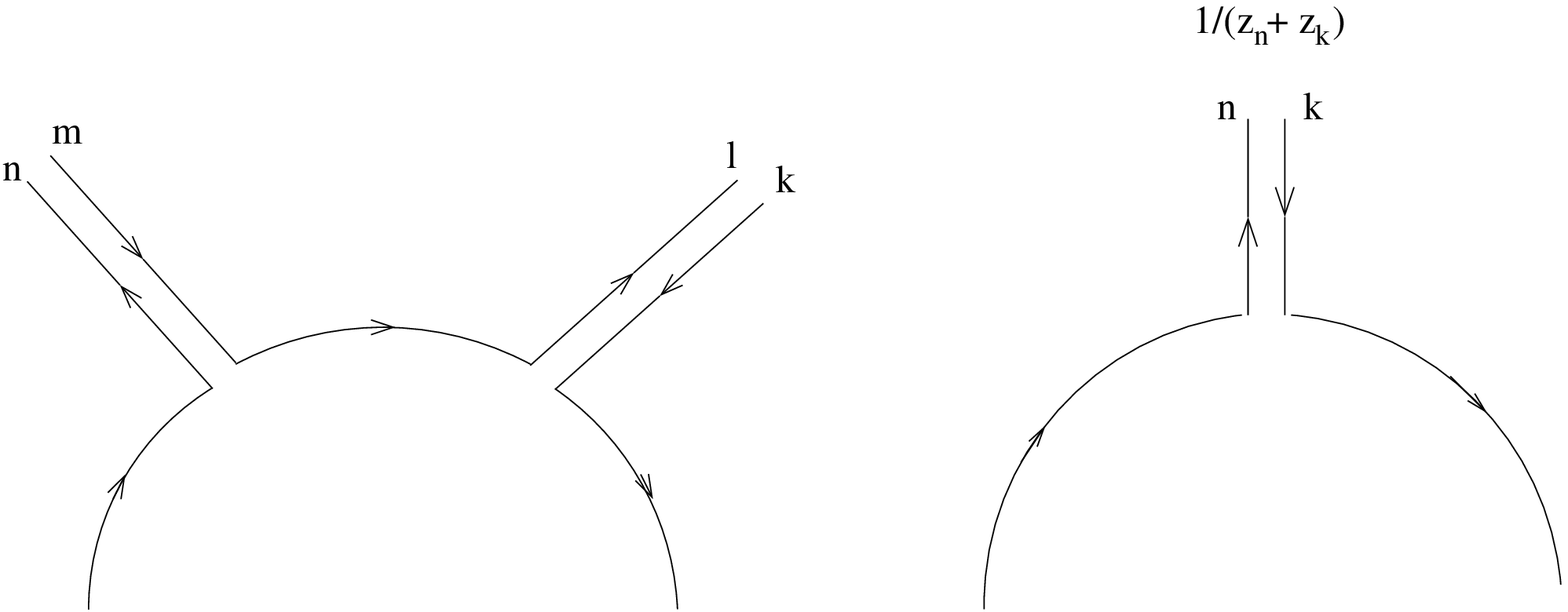}\\
{$\centerline{\rm (b)}$}
\end{center}
\caption[x] {\footnotesize Contacts when $X_{lk}$ collides with
$X_{nm}$ from either one side (a) or the other (b).}
\label{f:contacts.eps}
\end{figure}
By taking more $\phi$ derivatives of (\ref{recursion}) one obtains
the generalization of the recursion relation (\ref{twopoints}) 
for higher point functions. The case
of the 3-point function is somewhat special, due to the term
linear in $\phi$ that appears in the R.H.S. of Eq. (\ref{recursion}):
taking the derivative of this equation with respect to $\phi_{pq}$
gives
\be
\langle X_{lk}\, X_{qp}\,X_{nm}\rangle_{c.a.}
={D^{(2)}\over D\, Z_{kl}}
{\;\partial^2 H(Z,\phi)\;\over \partial \phi_{pq}\, \partial \phi_{mn}}-
{1\over g_s} (\delta_{lm}\,\delta_{np}\,\delta_{qk}+\delta_{lp}\,\delta_{qm}
\,\delta_{nk})
\label{threepoints}
\ee
where now the connection in the covariant derivative 
${D^{(2)}\over D\, Z_{lk}}$ with respect to $Z_{lk}$ 
includes the terms associated with the contacts between $X_{lk}$ with
two operators, $X_{nm}$ and $X_{qp}$:
\bea
&&{D^{(2)}\over D\, Z_{kl}}
{\;\partial^2 H(Z,\phi)\;\over \partial \phi_{pq}\, \partial \phi_{mn}}={\partial\over \partial\, Z_{kl}}
{\;\partial^2 H(Z,\phi)\;\over \partial \phi_{pq}\, \partial \phi_{mn}}-
{\delta_{lm}\over z_n+z_k}{\;\partial^2 H(Z,\phi)\;\over \partial \phi_{pq}\, \partial \phi_{kn}}-\nonumber\\
&&\qquad\qquad -{\delta_{nk}\over z_m+z_l}{\;\partial^2 H(Z,\phi)\;\over \partial \phi_{pq}\, \partial \phi_{ml}} -{\delta_{lp}\over z_q+z_k}{\;\partial^2 H(Z,\phi)\;\over \partial 
\phi_{kq}\, \partial \phi_{mn}}-\nonumber\\
&&\qquad\qquad\qquad -
{\delta_{qk}\over z_p+z_l}{\;\partial^2 H(Z,\phi)\;\over \partial \phi_{pl}\, \partial \phi_{mn}}
\label{covariantDtwo}
\eea
$n$-point functions of $X$ with $n>3$ are given by covariant $Z$-derivative
of the $n-1$-point functions:
\be
\langle X_{j_1i_1}\,X_{j_2i_2}\cdots X_{j_n i_n}\rangle_{c.a.} =
{D^{(n-1)}\over D\, Z_{i_1j_1}} \langle X_{j_2i_2}\cdots X_{j_n i_n}
\rangle_{c.a.}\qquad\quad {\rm for}\quad n>3
\label{npoints}
\ee
where the covariant derivative ${D^{(n-1)}\over D\, Z_{i_1j_1}}$ includes
the connection terms which correspond to the 
contacts of  $X_{j_1i_1}$ with the $n-1$ operators 
$X_{j_2i_2},\cdots, X_{j_n i_n}$.

Consistency of the recursion relations (\ref{npoints}) requires the 
Kontsevich covariant $Z$-derivative to be flat
\be
\Bigl[{D\over D\, Z_{kl}},{D\over D\, Z_{ij}}\Bigr] =0
\label{flatness}
\ee 
The flatness relation (\ref{flatness}) follows from the fact that the Kontsevich covariant $Z$-derivative 
can be rewritten as
\bea
\label{trivialization}
&&{D^{(n)}\over D\, Z_{kl}}\langle X_{j_1 i_1}\ldots X_{j_n i_n}\rangle_{c.a.}=
\Delta_{j_1 i_1;a_1 b_1}(Z)\ldots \Delta_{j_n i_n;a_n b_n}(Z)\times\\
&&\qquad\times\,{\partial \over \partial\, 
Z_{kl}}\Bigl[\Delta^{-1}_{a_1 b_1;c_1 d_1}(Z)\ldots \Delta^{-1}_{a_n b_n;c_n d_n}(Z) \langle X_{c_1d_1}\ldots
X_{c_nd_n} \rangle_{c.a.}\Bigr]\nonumber
\eea
and from the following symmetry of the Kontsevich propagator
\be
{\partial\over \partial\,Z_{i_1 j_1}}\,\Delta_{j_2 i_2;a b}(Z)=
{\partial\over \partial\,Z_{i_2 j_2}}\,\Delta_{j_1 i_1;a b}(Z)
\ee

\section{The disk} 

The tree approximation of the Kontsevich matrix theory computes disk
correlators of Liouville boundary cosmological constant operators. 
At tree level the one-point function vanishes
\be
\langle X_{nm}\rangle_{c.a.}^{tree} = 0
\ee
Therefore (\ref{twopoints}) and (\ref{threepoints}) imply that 
\be
{\;\partial^2 H^{tree}(Z,\phi)\;\over \partial \phi_{kl}\, \partial \phi_{mn}}
=0
\label{twopointstree}
\ee
and 
\be
\langle X_{lk}\, X_{qp}\,X_{nm}\rangle^{tree}_{c.a.}
=-
{1\over g_s} (\delta_{lm}\,\delta_{np}\,\delta_{qk}+\delta_{lp}\,\delta_{qm}
\,\delta_{nk})
\label{threepointstree}
\ee
Since this is $Z$-independent, it follows from the recursion relations
(\ref{npoints}) that the 
``bulk'' contributions to the string amplitudes with three or more
boundary cosmological constant operators vanish.
The only non-vanishing contributions to the
four and higher-point functions come from the Kontsevich connection.
From the point of view of the Liouville boundary conformal field theory
this means that the bulk contributions to 
string amplitudes with $n$ boundary cosmological
constant operators vanish when $n\ge 4$:
\be
\langle {\rm e}^{b\,\phi(x_1)}\cdots {\rm e}^{b\,\phi(x_n)}\rangle =0
\ee
Disk string amplitudes with $n\ge 4$  are given only by contact terms, 
coming from the boundary of the (open) string moduli space, 
i.e. from the regions where some of the points $x_i$ and $x_j$ collide.
This is the paradigmatic situation of topological string models. Typically,
contact contributions to string amplitudes are difficult to
compute from the world-sheet (first quantized) point of view: we have just
shown that the Feynman rules of Kontsevich open string field theory precisely
encode the information about such contacts, very much like Chern 
Simons field theory computes the contacts of open topological non-linear
sigma models \cite{W}. In the following of this Section we will discuss how
Kontsevich prediction for the disk amplitudes with $n$ 
boundary cosmological constant operators, contained in 
Eqs. (\ref{threepointstree}) and (\ref{npoints}), compares with the
computations \cite{SS}, \cite{KPS} of the same amplitudes which are based 
on the results about Liouville boundary conformal field theory 
on FZZT branes obtained in \cite{FZZ},\cite{T}.
 
Disk string amplitudes of boundary cosmological constant operators are precisely
the focus of the work  \cite{SS}, where it was observed that they are 
completely captured by the (closed) ground ring equations for the 
corresponding minimal string theory. 
The ground ring of the $(p,q)$ theory is generated by two elements, 
$x$ and $y$ subject to the relation
\be
T_p(y/C) -T_q(x)=0
\label{groundring}
\ee
where $T_n(x)$ are the Chebyshev polynomials of the first kind, and a $C$ is
a computable constant. Let $y= y_{p,q}(x)$ be the solution of the relation 
(\ref{groundring}), and define
the primitive $F_{p,q}(x)$ of $y_{p,q}(x)$:
\be
F_{p,q}(x) = \int^x\!\!\!\! dx'\,y_{p,q}(x')
\label{ssfreeenergy}
\ee
It was remarked in \cite{SS} that the disk string amplitudes with $n$ boundary
operators $\mathcal{B}_b(\phi)$ for the case of $N=1$ FZZT brane are given by the 
$n$-fold derivative of $F_{p,q}(x)$ with respect to $x$,
\be
\langle \mathcal{B}_b^n\rangle_{p,q} = {\partial^n\over\partial x^n} F_{p,q}(x)
\label{vbamplitudes}
\ee
upon the identification
\be
x = {\mu_B\over \sqrt{\mu}}
\label{ssmub}
\ee
where $\mu$ is the bulk Liouville cosmological constant\footnote{If $\mathcal{B}_b$ is the
vertex operator that multiplies $\mu_B$ in the Liouville action, its $n$-point amplitude
is actually given by Eq. (\ref{vbamplitudes}) times a dimensionfull factor 
$\sqrt{\mu}^{1/b^2 +1-n}$. In this section we set, for clarity, $\mu=1$: the $\mu$ dependence
can be easily reconstructed from dimensional arguments.}. 

Before comparing Liouville theory result (\ref{vbamplitudes}) for 
$(p,q)=(2,1)$ to the Kontsevich prediction let us first make a side remark.
Both $(p,q)$ and $(q,p)$ correspond to the same conformal theory, and
therefore the $(p,q)$ and the $(q,p)$ string theories should be identical. 
On the other hand the ground ring (\ref{groundring}) is invariant under the
exchange of $p$ with $q$ only if one simultaneously exchange $x$ with $y$.
It is therefore natural to consider, together with the $F_{p,q}(x)$, 
the primitive ${\tilde F}_{p,q}(y)$ of $x_{p,q}(y)$, 
the function inverse of $y_{p,q}(x)$:
\be
{\tilde F}_{p,q}(y) = \int^y\!\!\!\! dy'\,x_{p,q}(y')
\label{dualfreeenergy}
\ee
As pointed out in \cite{SS}, $F_{p,q}(x)$ and its ``dual'' 
${\tilde F}_{p,q}(y)$ are related by a Legendre transformation
\be
F_{p,q}(x) = x\,y -{\tilde F}_{p,q}(y)
\label{legendre}
\ee
Moreover, $y$ can be identified with the {\it dual} boundary cosmological
constant of Liouville theory
\be
y = {{\tilde \mu_B}\over \sqrt{{\tilde\mu}}}
\label{ssmubtilde}
\ee
where ${\tilde\mu}\equiv \mu^{1\over b^2}$ 
is the dual bulk cosmological constant. 

It is then clear that, because of the invariance of (\ref{groundring})
under the simultaneous exchange of $p$ with $q$ and $x$ with $y$, one has
\be
F_{p,q}(x) = {\tilde F}_{q,p}(x)
\label{simpledualityone}
\ee
and 
\be
{\tilde F}_{p,q}(y) = F_{q,p}(y)
\label{simpledualitytwo}
\ee
The identity (\ref{simpledualityone}) shows that one can compute the
disk string amplitude with $n$ boundary cosmological constant operators
either by differentiating $F_{p,q}(x)$ $n$ times with respect to the
boundary cosmological constant $\mu_B$ (at fixed $\mu$) or
differentiating the dual partition function ${\tilde F}_{q,p}(y)$ $n$ times
with respect to the dual boundary cosmological constant ${\tilde\mu}_B$
(at fixed ${\tilde \mu}$):
\be
\langle \mathcal{B}_b^n\rangle_{p,q} = {\partial^n\over\partial x^n} F_{p,q}(x)=
{\partial^n\over\partial y^n} {\tilde F}_{q,p}(y)\big|_{y=x}
\label{simpledualitythree}
\ee
Note, however, that  $F_{q,p}(x)\not= F_{p,q}(x)={\tilde F}_{q,p}(x)$:
one may, therefore, wonder what do the derivatives of ${\tilde F}_{p,q}=
F_{q,p}$ compute. Since $b^2=p/q$,  one might suspect that they compute 
n-point functions of the {\it dual} cosmological constant operator
$\mathcal{B}_{1\over b}(\phi)$
\be
\langle \mathcal{B}_{1\over b}^n\rangle_{p,q} {\mathop{=}^{?} }
{\partial^n\over\partial y^n} {\tilde F}_{p,q}(y)=
{\partial^n\over\partial x^n} F_{q,p}(x)\big|_{x=y}
\label{simpledualitywrong}
\ee
We are going to show that in fact this is not the case: actually if
(\ref{simpledualitywrong}) were true, one would face a puzzle. Indeed,
since $F_{q,p}\not= F_{p,q}$ no direct relation between the amplitudes
of the $(p,q)$ and the $(q,p)$ models would hold.  This would be surprising
since the two models should correspond to the same string theory.

Going back to the topological theory  $(p,q)=(2,1)$,  
the ground ring curve (\ref{groundring}) reduces in this case to
\be
y^2 -1-x=0
\label{groundringonetwo}
\ee 
Thus
\be
F_{2,1}(x) = \int^x\!\!\!\! dx'\,{1\over \sqrt{2}}\sqrt{x'+1} = 
{2\over 3}(x+1)^{3\over 2}
\label{ssfreeenergytwoone}
\ee
and
\be
{\tilde F}_{2,1}(y) = x(y)\, y -F_{2,1}(x(y)) ={1\over 3}\,y^3 -y
\label{ssfreeenergydualtwoone}
\ee
Thus, Liouville world-sheet computation for the
the string amplitude with $n$ boundary cosmological constant operators of
the $(2,1)$ theory gives
\be
\langle \mathcal{B}_b^n\rangle_{2,1} = {\partial^n\over\partial x^n}
{2\over 3}(x+1)^{3\over 2}
\label{npointstwoone}
\ee
Note however that had we chosen $(p,q)=(1,2)$ we would have obtained
\bea
&& \langle \mathcal{B}_b^3\rangle_{1,2} ={\partial^3\over\partial y^3}
F_{1,2}(y)={\partial^3\over\partial y^3}
{\tilde F}_{2,1}(y)= 2\nonumber\\
&&\langle \mathcal{B}_b^n\rangle_{1,2}={\partial^n\over\partial y^n} F_{1,2}(y)=
{\partial^n\over\partial y^n}{\tilde F}_{2,1}(y) =0\qquad {\rm for}\quad n\ge 4
\label{npointsonetwo}
\eea

To compare (\ref{npointstwoone}) and (\ref{npointsonetwo}) 
with Kontsevich prediction, let us specialize the general formula 
Eq. (\ref{npoints}) to the case of one single ($N=1$) brane on the disk
\be
\langle \mathcal{B}_b^n\rangle_{Kontsevich} = D_z^{(n-3)}\,
\langle \mathcal{B}_b^ 3\rangle_{Kontsevich} \qquad {\rm for}\quad n\ge 3
\label{npointskon}
\ee
where the covariant derivative is
\be
D_z^{(n-1)} = {\partial\over \partial z} - {n-1\over z}
\ee
Taking into account the value of the 
tree 3-point function (\ref{threepointstree})
\be
\langle \mathcal{B}_b^3\rangle_{Kontsevich} = -2
\label{threepointskon}
\ee
(having set for clarity $g_s=1$), one has
\be
\langle \mathcal{B}_b^n\rangle_{Kontsevich} = D_z^n F_K(z) \equiv D_z^{(n-1)}\,
D_z^{(n-2)}\,\cdots D_z^{(1)}\, D_z^{(0)}\, F_K(z)
\label{npointskonbis}
\ee
with
\be
F_K(z) = {2\over 3} z^3
\ee
In (\ref{npointskonbis}), $F_K(z)$ should be considered a scalar
function from the point of view of the covariant derivative, while its
$n-1$-th covariant derivative is a section of the $n-1$-th power
of the holomorphic tangent to $z$.

Because of the trivialization of the Kontsevich connection
(\ref{trivialization}), one can rewrite (\ref{npointskonbis}) as
\be
\langle \mathcal{B}_b^n\rangle_{Kontsevich}=
(\Delta(z))^n {\partial^n\over \partial x^n} F_K(z(x))
\ee
where $\Delta(z)= 2\,z$ is the Kontsevich inverse propagator and
$z(x)$ satisfies
\be
{\partial x\over \partial z} = \Delta(z) = 2\,z
\ee
This equation reproduces the ring relation (\ref{groundringonetwo}}) once we
identify the Kontsevich parameter $z$ with the dual boundary cosmological
constant $y$
\be
z=y
\label{kontoliouville}
\ee
With this identification,
\be
F_K(z(x)) = F_{2,1}(x)
\ee
and
\be
\langle \mathcal{B}_b^n\rangle_{Kontsevich} =
\Bigl({\partial x\over \partial y}\Bigr)^n \langle \mathcal{B}_b^n\rangle_{2,1}
\label{npointskonliou}
\ee
Finally, note that Kontsevich $n$-point amplitude can also be written as
follows
\be
\langle \mathcal{B}_b^n\rangle_{Kontsevich} = D_y^{n-3}\,\Bigl[-{\partial^3
{\tilde F}_{2,1}(y)\over \partial y^3}\Bigr]
\label{npointsdual}
\ee
Let us sum up what we learnt from the comparison of Kontsevich
and Liouville computations. 

First, it is clear that Kontsevich
theory really computes correlators of the $(1,2)$ minimal theory,
rather than those of the $(2,1)$ theory. This, in a sense, does
not come as a surprise since the authors of \cite{GR} did derive
Kontsevich theory from Witten open string field theories on Liouville
branes with $b^2=p/q = 1/2$ (rather than $b^2=2$). 

Moreover, Kontsevich theory shows that the string amplitudes
of the $(1,2)$ theory are {\it not} correctly computed in the Liouville
conformal field theory approach (\ref{npointsonetwo}), 
with the exception of the 3 point function. The reason of course
is that the conformal field theory computation misses the contact
terms which are instead correctly encoded in the Kontsevich connection.

Last --- and most important --- we understood the relation between
amplitudes of the $(2,1)$ and those of the $(1,2)$ theory: they are related
simply by a coordinate transformation on their common moduli space. 
Indeed, since the boundary cosmological constant operator is a (holomorphic)
vector tangent to the moduli space of FZZT branes, the $n$-point
amplitudes should transform as (holomorphic) tensors with $n$ indices.
This is precisely the content of Eq. (\ref{npointskonliou}). 
This is good news: after all the $(2,1)$ and the $(1,2)$ model should 
correspond to the same string theory. It is therefore comforting to verify
that the amplitudes of the two models do contain the same physical
information, and they merely parametrize the same moduli space with
different coordinates.  Introducing the Kontsevich connection also
nicely restores the symmetry between the partition function
$F_{1,2}(x)$ and (minus) its dual ${\tilde F}_{1,2}(y)$: one can start
with the 3-point functions computed with either partition functions,
and then compute higher-point functions with the appropriate
(covariant) derivative, using either Eq. (\ref{vbamplitudes}) or
Eq. (\ref{npointsdual})\footnote{It should be kept in mind that the
string amplitudes that are well defined on the disk are those with
$n\ge3$. The dual of (\ref{vbamplitudes}) is therefore
(\ref{npointsdual}), but {\it not} $-D_y^n {\tilde F}_{2,1}$.}.

The relation between the ground ring equation and the Kontsevich
model can be straightforwardly extended to the $(p,1)$ minimal
string models. For these theories the ground ring becomes
\be
T_p(y/C) -x=0
\label{p1ring}
\ee
Following the steps\footnote{
The computation that connects the curve (\ref{p1ring}) with
the Kontsevich model with potential (\ref{chebpotential})
is essentially the same performed in \cite{ADKMV}. 
That paper, however, deals with topological strings propagating in
a non-compact Calabi-Yau space. In that context, Eq. (\ref{p1ring}), 
rather than being interpreted as the ground ring equation of
non-critical bosonic strings, represents the locus where the
fibration which defines the Calabi-Yau degenerates.} 
discussed above, the corresponding Kontsevich model should 
be described by the action (\ref{kaction}) with a potential $V(M)$ 
given by
\be
V'(M) = T_p(M/C)
\label{chebpotential}
\ee
and the eigenvalues $z$ of the matrix $Z$ identified with the
dual boundary cosmological constants $z=y$ on the FZZT branes. 
One is lead therefore to conjecture that the Kontsevich model
with potential (\ref{chebpotential})
represents the (effective) open string field theory on the FZZT branes of the 
$(1,p)$ model: the Kontsevich field $X= M- Z$ should be identified with the 
boundary cosmological constant vertex operator $\mathcal{B}_b(\phi)$. 
Hopefully, this could
be proven along the lines of \cite{GR}, starting from the microscopic
Witten open string field theory. Even if the topological localization
of \cite{GR} would work for this case as well, the matrix theory
that one would obtain upon localization would still involve 
$q-1$ matrix fields.
One can further conjecture, following the suggestion of \cite{GR},
that, by integrating out all but the matrix
field corresponding to the vertex operator $\mathcal{B}_b(\phi)$, one obtains 
precisely the Kontsevich model (\ref{chebpotential}). 

Verifying directly these conjectures is a task that remains to be 
accomplished. One can however compare formulas
for the disk amplitudes of boundary cosmological constants that one 
obtains from the ground ring (\ref{p1ring}) with those predicted
by the Kontsevich model (\ref{chebpotential}). Our previous
discussion made clear that agreement between these two computations
has to hold only up to the appropriate 
coordinate transformation on the moduli space. 
Therefore the conjectured relation between amplitudes evaluated in the
two approaches should be just as in Eq. (\ref{npointskonliou}):
\be
\langle \mathcal{B}_b^n\rangle_{Kontsevich} =
\bigl[V_p''(y)\bigr]^n \langle \mathcal{B}_b^n\rangle_{p,1}
\label{npointskontopone}
\ee
where 
\be
V_p(y) = \int^y \!\!\!dy'\, T_p(y'/C)={\tilde F}_{p,1}(y)
\label{vp}
\ee
is the potential of the Kontsevich model\footnote{Note that $V_2(y)$ defined
by Eq.(\ref{vp}) is $V_2(y)=y^3/3-y$ (as in Eq. (\ref{ssfreeenergydualtwoone})), rather
than the purely cubic potential of Section 2. However, since in general
$F_K(z)=V(z)-z V'(z)$, the Kontsevich free energy $F_K(z)$ does not depend on 
the linear part of the potential.}. 
It is immediate to verify
that  (\ref{npointskontopone}) holds
if and only if the Kontsevich correlators for one single brane are given 
by covariant derivatives, that, when acting on correlators with $n$
vertex operators $\mathcal{B}_b(\phi)$, write as follows 
\be
D^{(n)}_y = \bigl[V_p''(y)\bigr]^n\,\partial_y\, {1\over 
\bigl[V_p''(y)\bigr]^n}=
\partial_y - n\,{V_p^{\prime\prime\prime}(y)\over V_p''(y)} 
\label{generalcovariant}
\ee
Following the steps of Section 2 one can indeed easily prove the validity
of Eq. (\ref{generalcovariant})
for the Kontsevich model with a generic potential $V(y)$ and $N=1$.  
The generalization of this computation to the case with $N$ branes
will be presented in the next Section,where, for sake of
clarity, we will restrict ourself to the case of monomial potentials
\be
V(M) = {M^{p+1}\over p+1}
\label{ppotential}
\ee
The case with a generic polynomial potential is not significantly
more complicated. 

Actually, it had been already suggested in the nineties 
\cite{KMMMZ},\cite{AM},\cite{GN},\cite{IZ} 
that the generalized Kontsevich models with monomial 
potentials  (\ref{ppotential}) are related to {\it closed} minimal strings of 
the  $(p,1)$ type with zero  bulk cosmological constant $\mu$. 
This also agrees with
the ground ring approach. In fact, taking into account the 
definitions (\ref{ssmub}) and (\ref{ssmubtilde}) of the 
parameters $x$ an $y$, one has that in the limit $\mu\to 0$ 
the ring (\ref{p1ring}) reduces to
\be
y^p - x=0
\label{p1ringmuzero}
\ee
in agreement with (\ref{ppotential}).

Let us conclude this Section by commenting on the implications of the
Kontsevich connection for the geometry of the moduli space of FZZT
branes.  From the point of the view of the conformal field theory,
this moduli space is parametrized by the boundary cosmological
constant $x$, which is the coupling constant of the marginal operator
$\mathcal{B}_b$. It is natural to analytically continue $x$ (and thus $y$) to
the complex plane.  Since at tree level $x$ and $y$ satisfy the ground
ring equation (\ref{p1ring}), the semi-classical moduli space is
identified with the genus zero Riemann surface $\mathcal{M}_{p,1}$
whose equation is (\ref{p1ring}). The uniformizing parameter for
$\mathcal{M}_{p,1}$ is $y$, i.e. the complex $y$ plane covers the
Riemann surface once and only once: We just understood that $y$ is the
same as $z$, the parameter that parametrizes the moduli space in the
Kontsevich open string field theory. Thus we notice two things: First,
as expected on general grounds \cite{SZ,S}, there is a non-trivial
reparametrization between the conformal field theory and the open
string field theory moduli, given by the relation $z=y(x)$ found in
(\ref{kontoliouville}). Moreover, in the case we are considering, 
string field theory provides a one-to-one parametrization 
of the quantum moduli space: This supports the proposal advanced in
\cite{SZ,S} that this might be 
a general property of Witten open string field theory.
   
Our analysis of Kontsevich string field theory also shows that 
on the quantum moduli space parametrized by 
$z$ plane there is a connection
\be
\Gamma_p=V_p''\,\partial_z (V_p'')^{-1}
\ee 
whose curvature\footnote{The curvature 
(\ref{curvature}) is associated with an holomorphic anomaly of
the Kontsevich string field theory.}
\be
R_p=\partial_{\bar z}\,\Gamma_p
\label{curvature}
\ee
has a delta function with support at the points where
\be
V_p''(z)=T'_p(z)=p\, U_{p-1}(z) = 0
\ee
($U_p(z)$ is the Chebyshev polynomial of the second kind). As $U_{p-1}(z)$ 
has degree $p-1$, the Kontsevich connection for $(p,1)$ minimal models defines 
a non-trivial line bundle of first Chern class equal to $p-1$. 

In this context, it is interesting to note that recently the authors
of \cite{MMSS} analyzed the structure of the quantum moduli space of
FZZT branes $\mathcal{M}_{p,q}$ in the general case $(p,q)$. At 
the semi-classical level $\mathcal{M}_{p,q}$ has singularities in the 
non-topological case $q>1$: these are singularities of the
{\it differential structure} on the moduli space which correspond 
to points $(x,y)$ on $\mathcal{M}_{p,q}$ that satisfy both 
\be 
U_{p-1}(y) = 0
\label{zzsingularitiesy}
\ee
and
\be 
U_{q-1}(x) = 0 
\label{zzsingularitiesx}
\ee
The physical interpretation of these singularities is related to ZZ branes.
It was found in \cite{MMSS} that these singularities disappear when 
non-perturbative effects are taken into
account. What we have found here is that in the $q=1$ case ---
when (\ref{zzsingularitiesx}) has no solutions --- the quantum
moduli space has {\it curvature} singularities located precisely
at the points that are solutions of (\ref{zzsingularitiesy}).
At these points in the moduli space the Kontsevich matrix theory
is singular. This means that the topological localization of \cite{GR}
fails precisely at these points of the moduli space: around these points
therefore the physics is described by the full fledged Witten open
string field theory and not by its Kontsevich reduction. It would
be interesting to understand the physical interpretation of this
phenomenon. 

\subsection{Geometrical origin of the Kontsevich connection}
In this subsection we will give a geometric interpretation of the 
Kontsevich connection, by showing that it is the Levi-Civita connection
associated with a (singular) metric on the moduli space.

We have seen that the curve in $C^{2}$ associated with the
Kontsevich model with potential $V(y)$ is
\be
V'(y) =x
\label{kcurve}
\ee
Consider the family of flat metrics on $C^2$ given by
\be
(d\, s)_\lambda^2 = d\,y\otimes d\,{\bar y}+\lambda^2 d\,x \otimes d\,{\bar x}
\ee
parametrized by $\lambda^2>0$. This metric induces the following metric
on the Kontsevich curve (\ref{kcurve})
\be
(d\, s)_{V,{\lambda}}^2 = \Bigl({1\over |V''(y(x))|^2} +\lambda^2 \Bigr) 
d\,x \otimes d\,{\bar x} =
\Bigl(1+\lambda^2\,|V''(y)|^2 \Bigr)\, d\,y \otimes d\,{\bar y}
\label{flatmetric}
\ee
Hence the associated Levi-Civita connection in local coordinates 
$(x,{\bar x})$ writes as follows
\be
\Gamma_{xx}^x = -{V^{\prime\prime\prime}(y(x))\over 
\Bigl(1 +\lambda^2\, |V''(y(x))|^2\Bigr)\,{V''}^2(y)}
\label{xconnection}
\ee
In local coordinates $(y,{\bar y})$ the connection is instead
\be
\Gamma_{yy}^y = { {\bar V}^{''}({\bar y})\, V^{\prime\prime\prime}(y)
\over {1\over \lambda^2}+ |V''(y)|^2}
\label{yconnection}
\ee
The limit of $\lambda\to\infty$ of such a connection is
\be
\Gamma_{xx}^x = 0
\label{xconnectioninfty}
\ee
in $x$ coordinates and
\be
\Gamma_{yy}^y = {V^{\prime\prime\prime}(y)\over V''(y) }
\label{yconnectioninfyt}
\ee
in $y$ coordinates. In conclusion the connection that appear in the
Kontsevich correlators is the connection induced by the flat 
metric on $C^2$ (\ref{flatmetric}) in the limit $\lambda\to\infty$.

\section{Generalizations to $q\not= 2$}
\subsection{$q>2$}
As recalled in Section 2, Kontsevich original model admits the following
generalization
when $q$ in (\ref{qpotential}) is greater then 2 
\be
{\rm e}^{F_q(g_s, Z)}= {\rm e}^{-{q\,\over (q+1)\, g_s} {\rm Tr}\,Z^{q+1}}
\int\! [dM] \, 
{\rm e}^{-{1\over g_s}{\rm Tr} \bigl[{M^{q+1}\over q+1} -Z^q\,M\bigr]}=
\int\! [dX] \, 
{\rm e}^{-{1\over g_s}\Gamma_{q+1}(X;Z)}
\label{kintegralq}
\ee 
where the action $\Gamma_{q+1}(X;Z)$ is the following invariant
polynomial of order $q+1$ in $X$:
\be
\Gamma_{q+1}(X;Z)\equiv {\rm Tr}{(X+Z)^{q+1}- Z^{q+1}- (q+1)\,
Z^q\, X\over q+1} 
\label{kactionq}
\ee
The quadratic part of the action (\ref{kactionq}) is
\be
\Gamma_{q}^{(2)} = 
{1\over 2}\sum_{ij,kl}\Delta_{lk;ji}^{(q)}(Z)\,X_{ji}\,X_{kl}
\ee
where we introduced the inverse propagator $\Delta_{lk;ji}^{(q)}(Z)$ of
the generalized Kontsevich theory
\be
\Delta_{lk;ji}^{(q)}(Z)=\sum_{a=0}^{q-1} Z^a_{lj}\,Z^{q-1-a}_{ik}
\ee
In the gauge in which $Z$ is diagonal the propagator becomes therefore
\be
\bigl[\Delta_{lk;ji}^{(q)}(Z)\bigr]^{-1} = {\delta_{ik}\,\delta_{jl}\over
P^{(q-1)}(z_i,z_j)}
\label{qpropagator}
\ee
where $P^{(q-1)}(z_i,z_j)$ is the homogeneous symmetric polynomial of
$z_i$ and $z_j$ of degree $q-1$
\be
P^{(q-1)}(z_i,z_j) \equiv \sum_{a=0}^{q-1} z_i^a\, z^{q-1-a}_j ={z_i^q-z_j^q\over z_i-z_j}
\label{qhomegenous}
\ee
Introduce the analog of the invariant function defined in Eq. (\ref{measure})
\be
{\rm e}^{f_q(g_s, Y)} = \int\! [dM] \,{\rm e}^{-{1\over g_s}{\rm Tr} 
\bigl[{M^q\over q+1} -Y\,M\bigr]}
\label{qmeasure}
\ee
Then the effective potential $H(Z,\phi)$ is given by
\bea
&&H(Z,\phi)= f_q(g_s,Y)- \log {\cal N}_q(Z)-{1\over g_s}\,
\Bigl[{q\over q+1} {\rm Tr}\,Z^{q+1}+ q {\rm Tr}\,Z^q\,\phi+
\nonumber\\
&&\qquad\qquad
+{1\over 2}{\rm Tr}\,\sum_{a=0}^{q-1} \phi\,Z^{a}\,\phi\,Z^{q-1-a} \Bigr]
\label{effectivepotentialq}
\eea
where
the matrix variable $Y_{ij}$ is
\be
Y_{ij} = (Z^{q})_{ij} + \sum_{a=0}^{q-1} (Z^a\,\phi\,Z^{q-1-a})_{ij}=
(Z^q)_{ij}+\phi_{kl}\,\Delta^{(q)}_{lk;ji}\equiv (Z^{q})_{ij} + L_Z(\phi)_{ij}
\ee
and $ L_Z(\phi)=0$ are the linearized equations of motion for $\phi$. In Eq. 
(\ref{effectivepotentialq}) we have also introduced the normalization factor
\be
{\cal N}_q(Z)=\int\! [dX] \,{\rm e}^{-{1\over g_s}\,\Gamma_{q}^{(2)}}=
({\rm Det}_{(lk;ji)}\, \Delta^{(q)}_{lk;ji})^{-1/2}
\ee
Thus
\be
{\partial f_q(Y)\over \partial Z_{kl}} =
{\partial f_q(Y)\over \partial \phi_{kl}}+{\partial f_q(Y)\over \partial 
\phi_{mn}}\,\bigl[\Delta^{(q)}(Z)\bigr]^{-1}_{nm;ji}\,
{\partial\over \partial Z_{kl}}\Delta^{(q)}_{qp;ji}(Z)\,\phi_{pq}
\label{phiderivativeq}
\ee
and the differential equation satisfied by $H(Z,\phi)$ becomes
\bea
&&\!\!\!\!\!\!\!\!\!\!\!
\Bigl[{\partial\over \partial Z_{kl}} - {\partial\over \partial \phi_{kl}}-
\Bigl(\bigl[\Delta^{(q)})(Z)\bigr]^{-1}_{nm;ji}\,
{\partial\over \partial Z_{kl}}\Delta^{(q)}_{qp;ji}(Z)\Bigr)\,\phi_{pq}{\partial\over \partial \phi_{mn}}
\Bigr] H(Z,\phi)=\\
&&\!\!\!\!\!\!\!\!\!\!\quad=-\Bigl[{\partial\over\partial Z_{kl}}-{\partial\over \partial \phi_{kl}}-\Bigl(\bigl[\Delta^{(q)}(Z)\bigr]^{-1}_{nm;ji}\,
{\partial\over \partial Z_{kl}}\Delta^{(q)}_{qp;ji}(Z)\Bigr)\,\phi_{pq}{\partial\over \partial \phi_{mn}}
\Bigr]\nonumber\\
&&\!\!\!\!\!\!\!\!\quad\Bigl[\log {\cal N}_q(Z)+{1\over g_s}\,\Bigl({q\over q+1} {\rm Tr}\,Z^{q+1}+ q {\rm Tr}\,Z^q\,\phi+
{1\over 2}{\rm Tr}\,\sum_{a=0}^{q-1} \phi\,Z^{a}\,\phi\,Z^{q-1-a}\Bigr)\Bigr]=\nonumber\\
&&\!\!\!\!\!\!={1\over 2}\bigl[\Delta^{(q)}(Z)\bigr]^{-1}_{qp;ji}\,
{\partial\over\partial Z_{kl}}\,\Delta^{(q)}_{ji;qp}(Z)+{1\over g_s}\,\sum_{a=0}^{q-1}\sum_{b=0}^{a-1} [Z^b\,\phi\,Z^{q-1-a}\,\phi\,Z^{a-1-b}]_{lk}
\label{qHidentity}
\eea
Consequently the covariant $\phi$-derivative is now
\bea
D_{lk}^{\phi}&\!\!\equiv\!\!&
{\partial\over \partial \phi_{kl}}+\Bigl(
\bigl[\Delta^{(q)}(Z)\bigr]^{-1}_{nm;ji}\,
{\partial\over \partial Z_{kl}}\Delta^{(q)}_{qp;ji}(Z)\Bigr)\,\phi_{pq}{\partial\over \partial \phi_{mn}}=\nonumber\\
&\!\!=\!\!& {\partial\over \partial \phi_{kl}}+
\sum_{ij} {1\over P^{(q-1)}(z_i,z_j) }\,
{\partial\over \partial Z_{kl}}\Bigl[\sum_{a=0}^{q-1} Z^{q-1-a}_{im}\,\phi_{mn}\, Z^{a}_{nj}\Bigl]{\partial\over \partial \phi_{ij}}
\eea
It follows, from the same argument as in section 2, that the connected amputated 2-point
function is the covariant $Z$-derivative of the 1-function
\bea
&&{\;\partial^2 H(Z,\phi)\;\over \partial \phi_{kl}\, \partial \phi_{mn}}=
\langle X_{lk}\,X_{nm}\rangle_{c.a.}-{1\over g_s}\Delta_{lk;nm}(Z) 
={D\over D\, Z_{kl}}\langle X_{nm}\rangle_{c.a.}
\label{twopointsq}
\eea
where the covariant $Z_{kl}$-derivative acts on $X_{nm}$ as
\bea
&&{D\over D\, Z_{kl}}\langle X_{nm}\rangle_{c.a.}=
{\partial\,\langle X_{nm}\rangle_{c.a.}\over \partial Z_{kl}} 
-\sum_{ij} {1\over P^{(q-1)}(z_i,z_j) }\times\nonumber\\
&&\qquad \times {\partial\over \partial Z_{kl}}\Bigl[\sum_{a=0}^{q-1} Z^{q-1-a}_{im}\, Z^{a}_{nj}\Bigl]
\langle X_{ji}\rangle_{c.a.}=\nonumber\\
&&\qquad = {\partial\,\langle X_{nm}\rangle_{c.a.}\over \partial Z_{kl}} 
-\sum_{ij} {1\over P^{(q-1)}(z_i,z_j) }\sum_{a=1}^{q-2}
\Bigl[{\partial (Z^a)_{im}
\over \partial Z_{kl}}Z^{q-1-a}_{nj} +\nonumber\\
&&\qquad\quad +{\partial (Z^a)_{nj}
\over \partial Z_{kl}}Z^{q-1-a}_{im}\Bigr]\langle X_{ji}\rangle_{c.a.}
\label{qZderivative}
\eea
Evaluating the $Z_{kl}$ derivatives in the gauge when $Z$ is diagonal, one
obtains
\be
{\partial (Z^a)_{im}\over \partial Z_{kl}}\Big|_{Z_{ij}=z_i\delta_{ij}} =
\delta_{ik}\,\delta_{ml}\, P^{(a-1)}(z_k,z_l)
\label{zaderivatives}
\ee
Thus, the covariant $Z$-derivative (\ref{qZderivative}) writes as follows
\bea
&&{D\over D\, Z_{kl}}\langle X_{nm}\rangle_{c.a.}=
{\partial\,\langle X_{nm}\rangle_{c.a.}\over \partial Z_{kl}} 
-\sum_{ij} {1\over P^{(q-1)}(z_i,z_j) }\times\nonumber\\
&&\qquad \times \sum_{a=1}^{q-2}
\Bigl[\delta_{ik}\,\delta_{ml}\,\delta_{nj}\,z_n^{q-1-a}\,
P^{(a-1)}(z_k,z_l) +\nonumber\\
&&\qquad\quad +\delta_{nk}\,\delta_{jl}\,\delta_{im}\,z_m^{q-1-a}\,
P^{(a-1)}(z_k,z_l)\Bigr]\langle X_{ji}\rangle_{c.a.}=\nonumber\\
&&\qquad =
{\partial\,\langle X_{nm}\rangle_{c.a.}\over \partial Z_{kl}} 
- {\delta_{ml}\,\sum_{a=1}^{q-2}
z_n^{q-1-a}\,P^{(a-1)}(z_k,z_l)\over P^{(q-1)}(z_k,z_n) }\langle X_{nk}\rangle_{c.a.}-\nonumber\\
&&\qquad - {\delta_{nk}\,\sum_{a=1}^{q-2} z_m^{q-1-a}\,
P^{(a-1)}(z_k,z_l)
\over P^{(q-1)}(z_m,z_l) }\langle X_{lm}\rangle_{c.a.} 
\label{qZderivativebis}
\eea
Now
\be
\sum_{a=1}^{q-2}
z_n^{q-1-a}\,P^{(a-1)}(z_k,z_l)= {P^{(q-1)}(z_l,z_n)- P^{(q-1)}(z_k,z_n)\over z_l-z_k}
\ee
Therefore
\bea
&&\!\!{D\langle X_{nm}\rangle_{c.a.}\over D\, Z_{kl}}=
{\partial\,\langle X_{nm}\rangle_{c.a.}\over \partial Z_{kl}} 
- \delta_{ml}{P^{(q-1)}(z_l,z_n)- P^{(q-1)}(z_k,z_n)\over (z_l-z_k)\, 
P^{(q-1)}(z_k,z_n)}\langle X_{nk}\rangle_{c.a.}-\nonumber\\
&&\qquad\quad -\delta_{nk}\,{P^{(q-1)}(z_l,z_m)- P^{(q-1)}(z_k,z_m)\over (z_l-z_k)\, P^{(q-1)}(z_m,z_l) }\langle X_{lm}\rangle_{c.a.} 
\label{qZderivativetris}
\eea
Thus, in the general case as well, the Kontsevich connection in
(\ref{qZderivativetris}) includes two terms that correspond, in the
first-quantized world-sheet picture, to the two contact terms that can
arise when the operators $X_{nm}$ and $X_{lk}$ collide, as shown in
Figure~\ref{f:contacts.eps}.  It is worth noting that in the $q>2$
case the factor associated with the contact that arises when an open
string stretching from the brane $z_l$ to the brane $z_k$ collides
with a string stretching from brane $z_k$ to a third brane $z_m$
depends not only on the final branes $z_l$ and $z_m$ but also on the
intermediate brane $z_k$.
 
Again one can verify that covariant $Z$-derivatives are flat
since they can be written as 
\bea
\label{qtrivialization}
&&{D^{(n)}\over D\, Z_{kl}}\langle X_{j_1 i_1}\ldots X_{j_n i_n}\rangle_{c.a.}=
\Delta^{(q)}_{j_1 i_1;a_1 b_1}(Z)\ldots \Delta^{(q)}_{j_n i_n;a_n b_n}(Z)\times\\
&&\qquad\times\,{\partial \over \partial\, 
Z_{kl}}\Bigl[\bigl[\Delta^{(q)}(Z)]^{-1}_{a_1 b_1;c_1 d_1}\ldots \bigl[\Delta^{(q)}(Z)\bigr]^{-1}_{a_n b_n;c_n d_n}\langle X_{c_1d_1}\ldots
X_{c_nd_n} \rangle_{c.a.}\Bigr]\nonumber
\eea
where $\bigl[\Delta^{(q)}(Z)\bigr]^{-1}_{ij;kl}$ is the propagator (\ref{qpropagator}) appropriate for the generalized Kontsevich model.

In the case of only one brane, $N=1$, the previous formulas considerably 
simplify. For the Kontsevich model with generic potential $V(M)$, the inverse
propagator is 
\be
\Delta^{(V)}(z)=V''(z)
\ee
and thus the covariant $Z$-derivative given in Eq. (\ref{qtrivialization}) 
reduces to
\be
D_z^{(n)}=\partial_z - n\,{V'''(z)\over V''(z)}
\ee 
which is the announced result (\ref{generalcovariant}). 

\subsection{$c=1$}
In \cite{IM} the Kontsevich matrix model for $c=1$ bosonic strings at 
self-dual radius was derived: it is defined by the following functional
integral
\be
Z_K(t,{\bar t})= \int\! [dM] \, {\rm e}^{-{\rm Tr}\,\bigl[(N-\nu)\,\log M +\nu
\,\sum_{k=1}^\infty {\bar t}_k \,M^k\bigr] +\nu {\rm Tr}\,M A}
\label{Kc=1}
\ee
where 
\be
\nu=-i\mu
\ee
with $\mu$ the bulk cosmological constant. $t_n$ and ${\bar t}_k$, which are 
the moduli corresponding to left and right moving closed string tachyons, 
enter the Kontsevich $c=1$ model (\ref{Kc=1}) in a quite 
asymmetric way: ${\bar t}_k$ are coupling constants of the invariant
operators ${\rm Tr}\,M^k$, while $t_n$ are expressed in terms of the 
external matrix source $A$ via the Frobenius-Miwa-Kontsevich transformation
\be
t_k={1\over n\nu}\,{\rm Tr}\,A^{-k}
\ee
The partition function $Z_K(t,{\bar t})$ encodes amplitudes between 
left moving and right moving closed string tachyon vertex operators 
$\mathcal{T}_k$ and $\mathcal{T}_{-k}$:
\be
Z_K(t,{\bar t})=({\rm det}\,A)^{-\nu}\,
\langle{\rm e}^{\sum_k t_k \mathcal{T}_k + 
{\bar t}_k \mathcal{T}_{-k}}\rangle_{closed}
\ee
Following the lesson of non-critical minimal strings, one would like to 
conjecture \cite{GR,M,GMM} that even the $c=1$ Kontsevich model should be 
identified with 
the (effective) open string field theory of $N$ stable branes of Liouville
theory coupled to $c=1$ matter. Moreover, as in the cases studied so far, we 
propose to identify the $X$ matrix field with the cosmological boundary vertex 
operator. In the following we want to derive the implications of this 
identification for amplitudes of cosmological boundary vertex operators 
at $c=1$. 

The problem of computing these amplitudes directly in the boundary conformal 
field theory is a subtle one \cite{MTV} and has not been solved yet: This is 
due to the fact that taking the naive $b\to 1$ limit of Liouville correlators 
produces divergent answers. A step towards the solution of this problem has 
been taken
in \cite{GMM}, where finite correlators of boundary tachyon operators of 
non-zero 
momentum has been obtained by a regularization and renormalization procedure.
However the naive extension of the correlators of \cite{GMM} to zero-momentum boundary operators --- which correspond to the
boundary cosmological constant operators we are interested in --- again produces infinite answers. It is possible that a further renormalization is needed to 
obtain a finite result. So, at the moment, the Kontsevich model for $c=1$ 
provides a prediction for boundary cosmological constant operator amplitudes that cannot
be independently verified.  

We will work at the ``topological'' point at which ${\bar t}_n=0$. 
In this case, by setting
\be
\nu\,Z=\nu'\,A^{-1}
\ee
with
\be
\nu'=\nu-N
\ee
the function integral (\ref{Kc=1}) reduces to one of the form 
(\ref{kfree}-\ref{kaction}) with potential
\be
V(M)=-\nu'\,\log M
\label{c=1potential}
\ee
Thus, the partition function of the $c=1$ Kontsevich model at the
topological point is
\be
{\rm e}^{F_{c=1}(\nu', Z)}= {\rm e}^{\nu'\,{\rm Tr}\,(1-\log Z)}
\int\! [dM] \, 
{\rm e}^{\nu' {\rm Tr} \bigl[\log M -Z^{-1}\,M\bigr]}=
\int\! [dX] \, 
{\rm e}^{-\nu' \Gamma_{c=1}(X;Z)}
\label{kintegralc=1}
\ee 
where $\Gamma_{c=1}(X;Z)$ is the Penner-like action
\be
\Gamma_{c=1}(X;Z)\equiv -{\rm Tr}\,\bigl[\log (1+Z^{-1} X) - Z^{-1} X\bigr]=
 \sum_{n=2}^\infty {(-1)^{n}\over n}\, {\rm Tr}\,(Z^{-1} X)^n
\label{kactionc=1}
\ee
The quadratic part of the action and the inverse propagator 
are
\be
\Gamma_{c=1}^{(2)} = {1\over 2}{\rm Tr}\,Z^{-1} X Z^{-1} X =
{1\over 2}\sum_{ij,kl}\Delta_{lk;ji}^{(c=1)}(Z)\,X_{ji}\,X_{kl}
\ee
\be
\Delta_{lk;ji}^{(c=1)}(Z)=Z^{-1}_{ik}\,Z^{-1}_{lj} 
\ee
The effective potential of the model is given by
\bea
&&H(Z,\phi)= f_{c=1}(\nu',Y)- \log {\cal N}_{c=1}(Z)-\nu'
\Bigl[{\rm Tr}\,(\log Z-1)+ \nonumber\\
&&\qquad\qquad+{\rm Tr}\,Z^{-1} \phi +
{1\over 2}{\rm Tr}\,Z^{-1} \phi Z^{-1} \phi \Bigr]\nonumber\\
&&{\rm e}^{f_{c=1}(\nu', Y)} = \int\! [dM] \,{\rm e}^{\nu' {\rm Tr} 
\bigl[\log M + Y\,M\bigr]}\nonumber\\
&&Y=\nu'\,Z^{-1}\,(-1+ \phi Z^{-1})\nonumber\\
&&{\cal N}_{c=1}(Z)=\int\! [dX] \,{\rm e}^{-\nu'\,\Gamma_{c=1}^{(2)}}
\label{effectivepotentialc=1}
\eea
and it satisfies the following differential equation 
\bea
&&\!\!\!\!\!\!\!\!\!\!\!\!\!\!\!\!\!\!\!\!\!\
\Bigl[{\partial\over \partial Z_{kl}} - {\partial\over \partial \phi_{kl}}+
\sum_i \Bigl(z_k^{-1}\,\phi_{ik}{\partial\over \partial \phi_{il}}+
z_l^{-1}\,\phi_{li}{\partial\over \partial \phi_{ki}}\Bigr)
\Bigr] H(Z,\phi)=\nonumber\\
&&=-N\,\delta_{kl} z^{-1}_k -(Z^{-1}\phi Z^{-1}\phi Z^{-1})_{lk}
\eea
In the equation above we have used the fact that the Kontsevich connection is,
in this case
\be
-\bigl[\Delta^{(c=1)}(Z)\bigr]^{-1}_{nm;ji}\,
{\partial\over \partial Z_{kl}}\Delta^{(c=1)}_{qp;ji}(Z)=
z^{-1}_k\,\delta_{kp}\,\delta_{lm}\,\delta_{nq} + z^{-1}_l\,\delta_{lq} 
\delta_{kn} \delta_{mp}
\ee
From Eq. (\ref{effectivepotentialc=1}) 
we can read off the covariant $Z$-derivative acting $X_{nm}$
\be
{D\over D\, Z_{kl}}\langle X_{nm}\rangle_{c.a.}=
{\partial\,\langle X_{nm}\rangle_{c.a.}\over \partial Z_{kl}} 
+z_l^{-1}\,\delta_{lm}\,\langle X_{nk}\rangle_{c.a.}+ z_k^{-1}\,\delta_{nk}\,\langle 
X_{lm}\rangle_{c.a.}
\label{covc=1}
\ee 

To make a prediction for amplitudes of boundary cosmological constant vertex operators 
on the disk we compute the Kontsevich effective potential $H(Z,\phi)$ at tree 
level. 
Since, in the tree approximation,
\be
f_{c=1}^{tree}(\nu', Y)=-\nu'\,{\rm Tr}\,(\log Y +1) 
\ee
we find, from Eq. (\ref{effectivepotentialc=1})
\be
\!\!H(Z,\phi)^{tree}= -\nu' {\rm Tr} \log (1-\phi Z^{-1}) - 
\log {\cal N}_{c=1}(Z)-\nu' \Bigl[{\rm Tr} Z^{-1} \phi +
{1\over 2}{\rm Tr} Z^{-1} \phi Z^{-1} \phi \Bigr]
\ee
By taking the third derivative with respect to $\phi$ of $H(Z,\phi)^{tree}$, we
deduce that the connected amputated 3-point function at tree level is
\be
C^{(3)}(z)=
\nu'\,{\delta_{il}\,
\delta_{kq}\,\delta_{jp}+ \delta_{iq}\, \delta_{kj}\,\delta_{lp}\over 
z_i\,z_k\,z_p}
\label{3pointc=1}
\ee
The two terms correspond to the two cyclically inequivalent orderings of the 
vertex operators. Higher point amplitudes are obtained from $C^{(3)}$ via the 
covariant derivative of Eq. (\ref{covc=1}):
\be
C^{(n)}(z)=D_z^{n-3}\,C^{(3)}(z)
\label{derc=1}
\ee
In particular, in the case of only one brane, the $n$-point amplitude is
\be
C^{(n)}(z)=\nu'\, {(n-1)!\over z^n}
\label{npointc=1}
\ee

As discussed above, the correlators $C^{(n)}$ should be identified with the amplitudes
of $n$ boundary cosmological constant operators on the disk. In order
to compare with boundary conformal field theory computations one needs
to know the relation between the string field theory parameter $z$ and the
conformal field theory parameter $x=\mu_B/\sqrt{\mu}$. It is natural to 
extend the recipe valid for the $(p,1)$ minimal models to the $c=1$ 
strings as well: as the analog of the curve (\ref{p1ring}) for the logarithmic 
potential (\ref{c=1potential}) is 
\be
xy=1
\label{c=1ring}
\ee
we propose the identification
\be
z=y=x^{-1}
\label{c=1z}
\ee
As we explained in section 3, the fact that amplitudes of boundary 
cosmological constant operators are given by the Kontsevich result (\ref{npointc=1}),
 with the identification (\ref{c=1z}), is equivalent to the statement 
that, for $c=1$ non-critical strings, these amplitudes are 
captured by the curve (\ref{c=1ring}) via the procedure of \cite{SS}. 

As a matter of fact, the curve  (\ref{c=1ring}) plays the same role 
for the topological B-model on the conifold that the curve (\ref{p1ringmuzero})
plays  for the $(p,1)$ topological models \cite{ADKMV} .  However, we note that  
the conjecture (\ref{c=1ring}), (\ref{c=1z}) appears to be somewhat  problematic in the non-topological formulation:
The curve (\ref{c=1ring}) coincides with the ground ring of non-critical
$c=1$ strings only in the case in which the scalar field representing the 
$c=1$ matter is non-compact. In the compact case, which is the one described by
the Kontsevich-like model (\ref{Kc=1}), the ground ring has two more 
generators, $u$ and $v$, and it is given by
\be
xy-uv=1
\label{c=1compactifiedring}
\ee
It is not clear to us how the extra generators find their place
in a suitable generalization --- if it exists --- of the mechanism discovered in \cite{SS} 
by which the closed ring computes open string amplitudes.

It is also not clear how the $b\to 1$ limit of Liouville correlators would
reproduce the prediction (\ref{npointc=1}). As recalled before, the naive
limit produce infinite expressions and an appropriate renormalization
procedure, for the case of boundary cosmological constant operators, has not yet been
defined.

\section{Conclusions}
We have shown that the interpretation of  Kontsevich matrix models as open string field
theory for FZZT branes coupled to minimal matter implies interesting  predictions
for amplitudes of boundary cosmological constant operators. We have learned that Kontsevich theory provides both global coordinates for the open string moduli space and a connection on it: 
By expressing insertions of boundary cosmological constant operators as covariant derivatives
with respect to this connection, Kontsevich theory captures contact term contributions
to open string amplitudes, arising from the points in moduli space in which boundary operators
collide either between themselves or with nodes of the Riemann surface.

We have shown that the consistency of Kontsevich  theory with boundary 
conformal field theory formulas hinges on some rather subtle issues. We have 
analyzed these by comparing Kontsevich amplitudes with
a representation of boundary cosmological constant correlators of $(p,1)$ 
non-critical strings on the disk recently found in \cite{SS}. 
In order to prove the agreement between the first and the second quantized
point of view two ingredients were essential:
The determination of the change of variables between the open string field
 theory and the boundary conformal field theory coordinates on the brane 
moduli space and the realization
that correlators of boundary operators transform, under such reparametrizations, as sections
of the tensor product of the (holomorphic) tangent bundle to the moduli space. 
The agreement we found provides evidence for the conjecture,  
which had not yet been explicitly verified for the case $p>2$, 
that generalized Kontsevich models do indeed represent (effective) 
open string field theories of non-critical 
bosonic strings coupled to $(p,1)$ minimal matter.

We have also presented an analysis of the Kontsevich model for $c=1$ non-critical strings along similar
lines. In the $c=1$ case, however, the current limited understanding of Liouville theory in the $b\to 1$ 
limit did not allow us to compare  Kontsevich predictions with boundary conformal field
theory results. It would, of course, be interesting to elucidate this issue. One possibility
is that, once an appropriate renormalization procedure at $b\to 1$ has been taken, amplitudes
of boundary cosmological constant operators do indeed reduce to Kontsevich correlators. Another
possibility is that the Kontsevich matrix field $X$ is not, in this case, the string field corresponding
to the cosmological constant operator but some more complicated composite field, originating
from the process of integrating out all other fields in the full string field theory.

Perhaps one of the most intriguing results of our analysis is the fact that the Kontsevich connection
is not flat and its curvature has delta function singularities at the points where the Kontsevich
kinetic term degenerates. The non-flatness of the connection is associated with a holomorphic anomaly
of the Kontsevich matrix model, analogous to the one of closed topological 
strings \cite{BCOV}. The holomorphic anomaly of closed topological strings has deep consequences for the background independence of the models \cite{W2}. It would be interesting to see if a similar interpretation 
of the anomaly could be given in the Kontsevich case also.
One should also elucidate the physical meaning of the singularities of the Kontsevich curvature. 
It has recently been shown in \cite{MMSS} that the quantum 
moduli space of open minimal string theory is smooth, as a differentiable
manifold. The poles of the Kontsevich connection imply that the same
moduli space, however,  has curvature singularities. At the singularities
topological localization fails and one presumably should reintroduce the full degrees 
of freedom of Witten open string field theory to correctly describe the physics around such points.

\section*{Acknowledgments} 

We thank I. Kostov, B. Ponsot, L. Rastelli, N. Seiberg, J. Teschner
and A. Zaffaroni for useful discussions and correspondence. 
This work was supported in part by
the European Commission's Human Potential program under contract
HPRN-CT-2000-00131 Quantum Space-Time, to which C.I. is
associated through the Frascati National Laboratory. S.G. was supported
by an I.N.F.N. fellowship.

\end{document}